\documentclass[12pt]{iopart}
\usepackage{graphicx}
\usepackage{epsfig}
\usepackage{iopams}
\begin{document}

\title{Anomalous diffusion in disordered multi-channel systems}
\author{R\'obert Juh\'asz}
\address{Research Institute for Solid
State Physics and Optics, H-1525 Budapest, P.O.Box 49, Hungary}
\ead{juhasz@szfki.hu}
\author{Ferenc Igl\'oi}
\address{Research Institute for Solid
State Physics and Optics, H-1525 Budapest, P.O.Box 49, Hungary}
\address{
Institute of Theoretical Physics,
Szeged University, H-6720 Szeged, Hungary}
\ead{igloi@szfki.hu}

\begin{abstract}
We study diffusion of a particle in a system composed of $K$ parallel channels,
where the transition rates within the channels are quenched random variables
whereas the inter-channel transition rate $v$ is homogeneous. 
A variant of the strong disorder renormalization group method and Monte Carlo
simulations are used. Generally, we
observe anomalous diffusion, where the average distance travelled by the particle, $[\langle x(t) \rangle]_{\rm av}$,
has a power-law time-dependence $[\langle x(t) \rangle]_{\rm av} \sim t^{\mu_K(v)}$,
with a diffusion exponent $0 \le \mu_K(v) \le 1$.  
In the presence of left-right symmetry of the distribution of random 
rates the recurrent point of
the multi-channel system is independent of $K$, and the diffusion exponent is found to increase with
$K$ and decrease with $v$. In
the absence of this symmetry, the recurrent point may be shifted 
with $K$ and the current can be reversed by
varying the lane change rate $v$.
\end{abstract}

\maketitle

\newcommand{\bc}{\begin{center}}
\newcommand{\ec}{\end{center}}
\newcommand{\be}{\begin{equation}}
\newcommand{\ee}{\end{equation}}
\newcommand{\beqn}{\begin{eqnarray}}
\newcommand{\eeqn}{\end{eqnarray}}

\section{Introduction}
Brownian motion is one of the most studied and the best understood stochastic
processes,
both in homogeneous and in random environments \cite{ab,havlin,haus,bouchaud}. 
Physical motivation to study diffusion in disordered media is
provided by transport processes (molecular diffusion, electric conduction, biological transport in a cell) on the one hand
and by relaxation in complex systems (in spin glasses and random ferromagnets) on the other. For asymmetric hoppings,
when the microscopic transition rates depend on the direction, too, disorder can strongly modify the properties of the transport
in dimensions, $d<2$ \cite{bouchaud}.

The effect of disorder is particularly strong in $1d$, in which case several
exact results are known \cite{solomon,kesten,sinai,golosov,derrida,zeitouni}. 
Many of those are
obtained in a mathematically rigorous way, such as Sinai-scaling of the
average square displacement, $[\langle x^2(t) \rangle]_{\rm av}$ in time, $t$ \cite{sinai}.
This behaves at the recurrent point as:
\be
[\langle x^2(t) \rangle]_{\rm av} \sim \ln^4(t)\;,
\label{sinai}
\ee
where $\langle \dots \rangle$ stands for the thermal average for a given
realization of disorder and $[\dots]_{\rm av}$ is used to denote averaging over quenched disorder.
Some presumably exact results are also calculated by a physically motivated method using a strong
disorder renormalization group approach \cite{dmf}. In this method sites of the lattice having the smallest barrier in the random potential landscape are
consecutively eliminated and new transition rates are perturbatively calculated between remaining sites. Thus during renormalization the
fastest rates are eliminated and at the fixed point only the slow excitations remain, which govern the cooperative dynamics of the
system. The fixed point of the transformation is a so called infinite disorder fixed point, at which the ratio of transition rates between
neighbouring sites goes either to zero or to infinity, so that the transformation becomes asymptotically exact. The infinite
disorder fixed point of the $1d$ random random walk is isomorphic with the fixed point of some $1d$ random quantum magnets\cite{mdh},
such as the random transverse-field Ising spin chain\cite{fisher} and some critical
properties of the two models are interrelated \cite{im}.

It is convenient to introduce a control parameter of the random random walk, $\delta$, which has the value $\delta=0$ at the recurrent point, which
corresponds to the fixed point of the strong disorder renormalization group transformation. In the recurrent point the correlation length of the system is
divergent. On the contrary for $\delta \ne 0$ when the walk is drifted to
one or another direction the correlation length is finite and this regime is called the transient phase.
In the transient phase the average travelled distance, $[\langle x(t) \rangle ]_{\rm av}$, is non-zero and has a power-low
time-dependence:
\be
[\langle x(t) \rangle ]_{\rm av} \sim t^{\mu_1}\;,
\label{x(t)}
\ee
where the diffusion exponent, $\mu_1=\mu_1(\delta) \le 1$, is a continuous function of the control parameter. According to renormalization group analysis
in the vicinity of the critical point it behaves as\cite{dmf,im}:
\be
\mu_1(\delta)=\delta+O(\delta^2)\;.
\label{mu1_delta}
\ee

In reality, the channel in which the transport takes place is often 
not strictly one-dimensional. For example, we may think of vehicular 
traffic in multi-lane roads \cite{santen} 
or the active transport in cells, which is
realized by molecular motors moving on filaments that are composed of several
proto-filaments \cite{howard}.  
Besides, the problem of one dimensional random walks with long (but bounded) 
jump lengths  is also equivalent to random walks with nearest neighbour jumps
in a strip of finite width. 
This problem arises also in the context of disordered iterated 
maps \cite{radons} or for random walks with inertia \cite{szta}. 
Comparatively, less results are known for random walks on strips \cite{key,bolthausen}, 
which can be considered to be in a way
between dimensions one and two. Only very recently, it has been shown 
that at the recurrent point of the multi-channel system, 
Sinai scaling in Eq.(\ref{sinai})
is valid \cite{bolthausen2}. 
The same conclusion is obtained by one of us by using a variant of the strong
disorder renormalization group method \cite{juhasz}. 
Indeed, during renormalization
the strip is renormalized to a random chain, thus results quoted above in Eqs. (\ref{x(t)}) and (\ref{mu1_delta}) are expected to hold in this case, too.

In this work we are going to study the properties of a random walk in a
disordered multi-channel system, we pay particular attention to the transient
phase. 
In this study
we consider systems composed of $K$ channels with quenched random transition 
rates within channels and allow symmetric transitions between channels 
with a site-independent lane change rate $v$. 
We measure the diffusion exponent in the system, $\mu_K(v)$, and study its
dependence on $K$ and $v$.
In the strong coupling limit, $v\to\infty$, and in the vicinity of the recurrent point we obtain analytical results. In the general situation we use numerical methods, either
Monte-Carlo simulations or numerical application of the strong disorder renormalization group method.

The rest of the paper is organized as follows. The model is presented in Sec. \ref{sec:model} and its properties are analyzed
in the strong coupling limit in Sec.\ref{sec:strong}. The essence of the
renormalization procedure and phenomenological scaling considerations are given
in Sec.\ref{sec:RG}. Numerical analysis for $K=2$ and for varying lane change rate
is given in Sec.\ref{sec:numerical}. Our results are discussed in the final Section.

\section{Model}
\label{sec:model}

\subsection{$1d$ systems}
We define the model first in one-dimension, on 
the discrete points, $l=1,2,\dots,L$, with periodic boundaries. 
A continuous time stochastic process is considered in which the particle 
jumps from site $l$ 
to site $l+1$ with a transition rate, $p_l$, which may be different from the transition rate of the reverse jump, $(l+1 \to l)$,
denoted by $q_{l+1}$. 
Here the pairs of rates ($p_l$,$q_l$) are independent 
and identically distributed random variables drawn from 
the distribution $P(p,q)$. 
We generally consider the limit $L\to\infty$ and are interested in 
the asymptotic (long time) behavior.

The control parameter of the $1d$ random walk is defined as\cite{dmf,im}:
\be
\delta=\frac{[\ln p]_{\rm av}-[\ln q]_{\rm av}}{{\rm var}[\ln p]+{\rm var}[\ln q]}\;,
\label{delta}
\ee
where ${\rm var}[y]$ stands for the variance of $y$. In the $1d$ problem, 
the mean velocity and the diffusion constant of a particle on a finite ring is 
expressed as a Kesten random
variable and therefore the diffusion exponent, $\mu_1$, can be calculated
exactly even for not small $\delta$ \cite{solomon,kesten,derrida}. 
It is given by the positive root of the equation:
\be
\left[\left( \frac{q}{p}\right)^{\mu_1} \right]_{\rm av}=1,
\label{mu1}
\ee
provided it is smaller than one. Otherwise the particle moves with a finite
mean velocity and $\mu_1=1$. 
Indeed for small $\delta$ Eq(\ref{mu1}) gives back the leading behavior of the renormalization group result in Eq.(\ref{mu1_delta}).

In this paper we consider two types of random environments. For {\it type A (or symmetric) randomness} the transition rates satisfy the relation that
the product, $p_l q_l$, does not depend on the position and the distribution
of the rates is symmetric at the recurrent point, i.e. $P(p,q)=P(q,p)$.
Here we use a bimodal disorder:
%
%
\be
p_l=\frac{1}{q_l}= \left\{
\begin{array}{c}
{ \lambda \qquad {\rm with ~probability} ~c, }\\
{ 1/\lambda \qquad {\rm with ~probability} ~1-c,}  
\end{array}
\right. 
\label{bim_A} 
\ee
for which the control parameter is given by:
\be
\delta_A=\frac{2c-1}{4c(1-c)} \frac{1}{\ln \lambda}\;,
\label{delta_A}
\ee
and the diffusion exponent is:
\be
\mu_1(A)=\left|\frac{\ln(c^{-1}-1)}{2 \ln \lambda}\right|\;.
\label{mu_1_A}
\ee
For {\it type B (or asymmetric) randomness} the distribution of rates 
is non-symmetric at the recurrent point either.
We have used the following distribution:
%
%
\be
p_l=1,\quad q_l= \left\{
\begin{array}{c}
{ \lambda \qquad {\rm with ~probability} ~c, }\\
{ 1/\lambda \qquad {\rm with ~probability} ~1-c,}  
\end{array}
\right. 
\label{bim_B} 
\ee
the control parameter is given by:
\be
\delta_B=-\frac{2c-1}{4c(1-c)} \frac{1}{\ln \lambda}\;,
\label{delta_B}
\ee
and the diffusion exponent is:
\be
\mu_1(B)=\left|\frac{\ln(c^{-1}-1)}{ \ln \lambda}\right|\;.
\label{mu_1_B}
\ee
\subsection{Multi-channel systems}
The multi-channel system consists of $i=1,2,\dots,K$ channels in which the
transition rates are drawn independently from each other but from the same
distribution $P(p,q)$.
Lane changes occur with rate $v$ between nearest neighbor lanes (also between lane $i=K$ and $i=1$), which does not depend on the position.
This means that the transition rate, $w[(l,i) \to (k,j)]$, for a jump from site $(l,i)$ to site $(k,j)$ is given by:
\be
w[(l,i) \to (k,j)]=\delta_{i,j}\left(p^{(i)}_l \delta_{l+1,k}+q^{(i)}_{l}\delta_{l,k}\right)+\delta_{l,k}\left(\delta_{i+1,j}+\delta_{i-1,j}\right)v\;.
\label{w}
\ee

Similarly to the one dimensional model, the multi-channel system has a 
recurrent point at, otherwise it is transient in the positive or negative
directions. The transient regions consist of two phases: in the sublinear
transient phase $\mu_K<1$, while in the ballistic phase $\mu_K=1$.

For symmetric randomness, it is easy to see that the recurrent point 
is not shifted 
for $K>1$ compared to that of $K=1$. 
It is due to the fact that
here recurrence is connected with the left-right symmetry of the distribution of the transition rates, which stays invariant after connecting the channels, too. On the
contrary, for asymmetric randomness, the condition of recurrence depends on $K$
and $v$.

\section{Strong-coupling limit}
\label{sec:strong}

First, we discuss the strong coupling limit ($v\to\infty$), 
which is simple enough to obtain analytical results.
In the following, the set of $K$ sites with the horizontal 
coordinate $l$ will be called the $l$th layer.  
In the limit $v \to \infty$, the particle being in the $l$th layer,
visits all sites of the layer infinitely many times 
before jumping to another layer. 
Consequently, the effective transition rate between layers is just the 
average of the interlayer jump rates over the vertical coordinate:
\be
\tilde{p}_l=\frac{\sum_i p^{(i)}_l}{K},\quad \tilde{q}_l=\frac{\sum_i q^{(i)}_l}{K}\;,
\label{eff_pq}
\ee
and the problem is reduced to an effective one-dimensional random walk. 
The corresponding $1d$ random environment is given by the
transition rates, $\tilde{p}_l$ and $\tilde{q}_l$,
the distributions of which follow from the distributions of the 
original chain, respectively. In the following we analyse the
properties of the strong coupling limit for the two different 
types of randomness.

\subsection{Symmetric randomness}

Let us consider a general symmetric randomness, where $p^{(i)}_lq^{(i)}_l=1$ 
at all $(l,i)$ sites and the probability density of $p^{(i)}_l$
is denoted by $\rho(p)$. 
As we have already mentioned, if $\delta_A=0$ then the system is
recurrent for any $K>1$, due to the symmetry of the distribution $P(p,q)$.

Specially for $K=2$, the diffusion exponent $\mu_2$ can be given in terms of 
$\mu_1$ as follows.
For the effective one-channel system with rates given in
Eq. (\ref{eff_pq})), equation (\ref{mu1}) can be written as:
\beqn
\int {\rm d} p_1 \rho(p_1) \int {\rm d} p_2 \rho(p_2) \left(\frac{1/p_1+1/p_2}{p_1+p_2} \right)^{\mu_2}=\nonumber \\
\int {\rm d} p_1 \rho(p_1) (p_1)^{-\mu_2} \int {\rm d} p_2 \rho(p_2) (p_2)^{-\mu_2}.
\eeqn
The diffusion exponent $\mu_1$ for the case $K=1$ is determined by 
\be
\int {\rm d} p \rho(p) (p)^{-2\mu_1}=1\;.
\ee
Comparing the latter two equations, we obtain the simple relation:
\be
\mu_2=2 \mu_1,\quad v\to\infty \;.
\label{mu2}
\ee
For $K>2$ a similar relation holds only for {\it weak disorder} and in the vicinity of the recurrent point. Here
for weak disorder we use the parametrization $p_l=1+\epsilon_l$, with $|\epsilon_l| \ll 1$ in which case in the $K$-channel system the control parameter, $\delta_A^{(K)}$
is simply related to the control parameter in the one lane system, $\delta_A$,
as $\delta_A^{(K)}\simeq K\delta_A$. Consequently one obtains for the
diffusion exponent from eq.(\ref{mu1_delta}):
\be
\mu_K \simeq K \mu_1,\quad \mu_1 \ll 1 \quad {\rm and}\quad v\to\infty\;.
\ee
Thus with increasing number of channels, the extension of the sub-linear transient regime, measured by $\delta_A$ is decreasing and, for any fixed $\delta_A>0$ 
and sufficiently large $K$, the transport becomes ballistic.

\begin{figure}[htb]
\begin{center}
\includegraphics[width=6.3cm,angle=0]{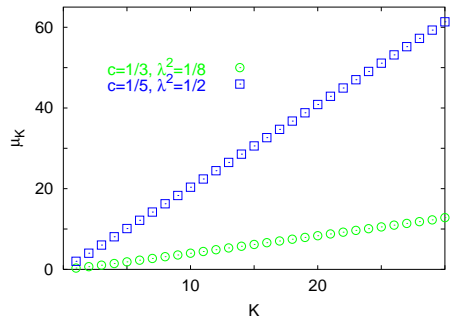}
\caption{Diffusion exponent, $\mu_K$, plotted against the number of channels, $K$, in the large coupling limit
for symmetric randomness in Eq.(\ref{bim_A}) calculated from Eq.(\ref{muK}).}
\label{fig1}
\end{center}
\end{figure}

For large values of $K$ the effective transition rates in Eq.(\ref{eff_pq}) will approach a Gaussian distribution with a relative mean deviation, which is decreasing with $K$.
Consequently the disorder of the effective couplings will be weak for large $K$, thus one expects that $\mu_K/K \to const$. We have checked this expectation
by using the bimodal distribution in Eq.(\ref{bim_A}), when for general $K$,
the diffusion exponent has been obtained by solving numerically the equation:
\be
\sum_{n=0}^K \frac{K!}{n!(K-n)!} c^{K-n} (1-c)^{n} \left(\frac{n \lambda^2 + K-n}{n+(K-n) \lambda^2} \right)^{\mu_K}=1\;.
\label{muK}
\ee
The obtained values of $\mu_K$ for different number of channels are shown in Fig.\ref{fig1} for two different bimodal parameters.
Indeed, $\mu_K$ is found to be approximately linear in $K$ 
for large $K$ in both cases.

\subsection{Asymmetric randomness}

Next, we consider the bimodal disorder in Eq.(\ref{bim_B}).
The probability $c=c^*_K$ for which the system is recurrent is the root of the 
following equation:
\be
\sum_{n=0}^K \frac{K!}{n!(K-n)!} (c^*_K)^{K-n} (1-c^*_K)^{n} \ln \frac{(K-n) \lambda + n \lambda^{-1}}{K}=0\;.
\ee
For $K=2$ it is explicitly given as:
\be
c^*_2(\lambda)=\frac{1}{2}(1+f-\sqrt{1+f^2}),\quad f=\ln \lambda / \ln[(\lambda+\lambda^{-1})/2]\;.
\ee
For weak disorder we have $\lim_{\lambda \to 1} c^*_2(\lambda)=c^*_1=1/2$,
which is shifted to $\lim_{\lambda \to \infty} c^*_2(\lambda)=1-\sqrt{2}/2$
in the strong disorder limit. For general $K$, this strong disorder limiting value is given by: $\lim_{\lambda \to \infty} c^*_K(\lambda)=1-2^{-1/K}$.
 
The dynamical exponent, $\mu_K$, is given as the positive root of the equation:
\be
\sum_{n=0}^K \frac{K!}{n!(K-n)!} (c)^{K-n} (1-c)^{n} \left(\frac{(K-n) \lambda + n \lambda^{-1}}{K}\right)^{\mu_K}=1\;.
\ee
It is easy to see that the boundary between the sub-linear and the
ballistic phase where $\mu_1=1$ is independent of $K$, 
since $\mu_K=1$ for any $K>1$ if $\mu_1=1$.

\section{Renormalization and phenomenological scaling}
\label{sec:RG}

\begin{figure}[htb]
\begin{center}
\includegraphics[width=6.3cm,angle=0]{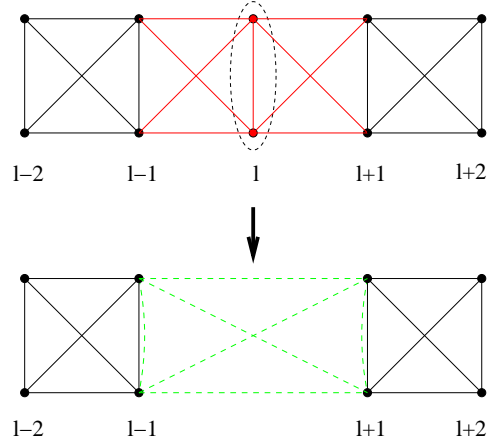}
\caption{Illustration of a renormalization step for $K=2$. Layer $l$ is decimated  and new transitions are created (indicated by dashed lines), the effective
  rates of which are calculated from the rates of transitions starting from or
  ending in the eliminated sites (shown in red).}  
\label{fig2}
\end{center}
\end{figure}

We are interested in the time
dependence of the displacement of the particle in the random environment
defined in the previous section. 
To obtain information on the dynamics in an infinite system, we shall
follow an indirect way in which 
finite systems of size $K\times L$ are considered 
and the dynamical exponent is estimated from the
finite-size scaling of the stationary probability current. 
For the estimation of the latter quantity we shall apply a real space
renormalization group method which is a variant of the strong disorder renormalization
group method and has
been used to study disordered quantum spin
chains \cite{mdh,fisher,im}, random walks in
one-dimensional random environments\cite{dmf} and random nonequilibrium processes\cite{contact,asep}. 
For higher dimensional dynamical systems, 
just as the present problem,  
the procedure has been formulated in terms of transition rates
 \cite{juhasz,monthus,vulpiani}:
states with the
largest exit rate are gradually removed from the configuration
space and one can infer the dynamics from the finite-size scaling
of the last remaining exit rate. This method has been applied to a
series of problems  \cite{juhasz,monthus,vulpiani}.  
A common difficulty of these methods in higher dimensions is that the
topology of the network of transitions between configurations
 does not remain invariant but
it becomes more and more complicated in the course of the procedure. 
In the present problem, one can however keep the topology invariant by 
eliminating complete layers, see Fig. \ref{fig2}.  
The rates of ``diagonal'' transitions 
$(l,i)\to (l\pm 1,j)$ with $i\neq j$ are initially zero (see Eq.(\ref{w})), but 
during the procedure they become positive. 
Thus, the invariant topology of the network of transitions for $K=2$ is a ladder
with both vertical and diagonal edges. 
In the followings, the precise definition of the renormalization 
method is given.   

In the steady state, the probability that the particle resides on site
$(l,i)$ will be denoted by $\pi_l(i)$. 
Introducing the row vectors
${\bf\pi}_l=(\pi_l(i))_{1\le i\le K}$, for a fixed random environment
(i.e. for a fixed set of transition probabilities),
the stationary probabilities satisfy the following 
system of linear equations which express the conservation
of probability\cite{juhasz}:
\be
{\bf \pi}_l{\bf S}_l={\bf \pi}_{l-1}{\bf P}_{l-1}+{\bf \pi}_{l+1}{\bf Q}_{l+1}, \quad 1\le l\le L.
\label{system}
\ee
Here, the $K\times K$ matrices ${\bf P}_l$,${\bf Q}_l$ and ${\bf S}_l$ are composed of the jump
rates as follows:
\beqn
P_l(i,j)=w[(l,i)\to (l+1,j)] \nonumber \\
Q_l(i,j)=w[(l,i)\to (l-1,j)] \nonumber \\ 
S_l(i,j)=-w[(l,i)\to (l,j)]  \quad i\neq j \nonumber \\ 
S_l(i,i)\equiv\sum_j[P_l(i,j)+Q_l(i,j)]-\sum_{j\neq i}S_l(i,j).
\eeqn
Let us introduce the exit rate of the $l$th layer as $\Omega_l:=1/\|{\bf S}_l^{-1}\|$, where the matrix
norm $\|\cdot\|$ is defined as $\|{\bf A}\|:=\max_{i}\sum_j|A(i,j)|$. 
In a renormalization step, layer $l$ is decimated, which
results in a one layer shorter system with effective rates in the adjacent 
layers given by 
\beqn
\tilde {\bf P}_{l-1}&=&{\bf P}_{l-1}{\bf S}_l^{-1}{\bf P}_l   
\nonumber \\
\tilde {\bf Q}_{l+1}&=&{\bf Q}_{l+1}{\bf S}_l^{-1}{\bf Q}_l    
\nonumber    \\
\tilde {\bf S}_{l-1}&=&{\bf S}_{l-1}-{\bf P}_{l-1}{\bf S}_l^{-1}{\bf Q}_l  
\nonumber    \\
\tilde {\bf S}_{l+1}&=&{\bf S}_{l+1}-{\bf Q}_{l+1}{\bf S}_l^{-1}{\bf P}_l.      \label{rule4}  
\eeqn
This transformation leaves the sojourn probability at the remaining sites, as
well as the probability current $J$ in the horizontal direction invariant.  
Following these rules, $L-1$ layers are decimated so that a single one 
remains. One can show that the remaining effective rates at the last remaining
layer are
independent of the order of elimination of the other $L-1$ layers. 
There are therefore only $L$ different states the procedure can end up
with, depending on the layer which is not decimated. 
The magnitude of the current that is invariant under the renormalization 
is related to the smallest among the $L$ possible last exit rates 
$\Omega_{\tilde l}\equiv\min_{l}\{\tilde \Omega_l\}$ as 
$|J|=|{\bf \pi}_{\tilde l}({\bf \tilde P}_{\tilde l}-{\bf \tilde Q}_{\tilde l})|\simeq\sum_i\pi_{\tilde l}(i)\Omega_{\tilde l}$ for large $L$. 

An efficient algorithm for finding the rate $\Omega_{\tilde l}$ 
is the following.
For the sake of convenience, let us assume that $L$ is an integer power of $2$.
First, the system is divided into two equal parts. The sites in the first part
are decimated (in an arbitrary order). This modifies the two surface layers of
the other part.
Then, starting again from the initial system, the second part is eliminated. 
This modifies the two surface layers of the first part.
This procedure is then applied recursively 
to each part with a halved length
until the procedure ends up with $L$ parts of length $1$. 
The exit rates of these $L$ layers are calculated and the smallest one is
selected.

The relation to strong disorder renormalization methods is more
transparent if the smallest remaining exit rate $\Omega_{\tilde l}$ 
is found in the way
that the layers with the actually largest exit rates are eliminated
one by one. Then the last remaining layer is expected to be layer 
$\tilde l$ with high probability.    
It has been shown that, in the course of the strong disorder 
renormalization, the vertical jump rates are non-decreasing while
the horizontal and diagonal jump rates tend to zero, so that 
the multi-channel system renormalizes to an effective one-channel
system \cite{juhasz}. 
Nevertheless, the previous algorithm needs only $O(L\ln L)$ operations 
per sample in contrast with the $O(L^2)$ operations of the second one. 
Another advantage of the former implementation is that it 
applies without difficulties also to discrete randomness, such as for
symmetric distribution where all exit rates are initially equal.     
In the numerical calculations we have therefore implemented 
the former algorithm.

In the following, the minimal exit rate $\Omega_{\tilde l}$ will be denoted by 
$\Omega_L$, as we are interested the scaling of this quantity with the system
size $L$ for fixed $K$.  
For a one-channel system ($K=1$) it is known from the analytical solution of
the renormalization equations that $\Omega_L$, in typical
samples, scales with the system size as $\Omega_L\sim L^{-1/\mu_1}$
in the transient phase, where $\mu_1$ is the unique positive root of
Eq. (\ref{mu1}) \cite{ijl}. 
On the other hand, from exact results on the time-dependence of displacement 
\cite{solomon,kesten,derrida}, 
one obtains that the current scales as $J\sim L^{-1/\mu_1}$ 
for $\mu_1<1$ and  $J\sim L^{-1}$ for $\mu_1>1$. Here,
$\mu_1$ is again the root of Eq. (\ref{mu1}).
Comparing the finite-size behavior of $J$ 
and $\Omega_L$
one concludes that the typical value of the probability at the last site 
$\tilde l$ must tend to an $L$-independent constant for large $L$ if $\mu_1<1$. 
This means that the walker typically spends a finite fraction of the time 
at a single site (favourite site). 
For $K>1$, we have no exact expression of $\mu_K$ at our disposal but  
the multi-channel system renormalizes to an effective 
one-channel system. Therefore we expect that the 
typical value of the probability of finding the particle in the 
favourite layer 
is still asymptotically $L$-independent. 
This implies that 
\be 
J\sim\Omega_L\sim L^{-1/\mu_K},
\label{JOmega}
\ee
where the first proportionality 
holds if the diffusion exponent $\mu_K$ is less than one.   
The importance of this relation is that the
diffusion exponent can be estimated by computing $\Omega_L$ which is
a much less numerical effort than the direct solution of Eqs. (\ref{system}).

In the frame of a phenomenological random trap picture of the system,
the renormalization procedure can be naturally interpreted. 
The random environment in which the particle moves can be thought of 
as a collection of consecutive, spatially localized trapping regions. 
In some regions, the particle may spend very
long times and obviously the trapping times in these regions determine
predominantly the traveling time of the particle. 
The renormalization procedure means a kind of coarse-graining of the
environment in the course of which trapping regions are step by step 
contracted and, simultaneously, trapping regions that had been
contracted to a single site are completely eliminated one by one 
in the order of increasing trapping time.
For late times, when the latter process dominates, 
the inverse of the effective exit rate gives roughly the release 
time from the corresponding trapping region.
The inverse of the last exit rate $\Omega_{L}$ can thus be interpreted as 
the mean release time from the ``deepest'' trap of the environment.  
Close to the fixed-point of the transformation (i.e. when the largest
effective exit rate is very small) the system can be treated as an effective
one-channel system. 
From the analytical treatment of the renormalization group
equations for $K=1$ it is known\cite{ijl} that the distribution of exit rates has a 
power-law tail $P_>(\tilde\Omega)\sim \tilde\Omega^{-\mu_K}$, which is expected to hold for $K>1$ (see
the numerical results in Fig. \ref{fig5}).
The mean time needed for the particle to make a complete tour on a 
finite ring is
given approximately as $t=1/J\sim \sum_i1/\tilde\Omega_i$. 
If $\mu_K<1$ (i.e. in the zero-velocity phase), the above sum of random
variables is
dominated by the largest one, i.e. $\sum_i1/\tilde\Omega_i\sim 1/\Omega_{L}$ 
and we obtain $J\sim \Omega_{L}$. 
We see, that this simple phenomenological theory 
corroborates relation (\ref{JOmega}).

\section{Numerical Results}
\label{sec:numerical}

We have applied two numerical methods for the investigation of the model:
Monte Carlo simulations and the renormalization group method. 
In case of Monte Carlo simulations, 
we have generated $10^4$ independent random environments with $L=10^7$, $K=2$ 
and simulated the process in each sample with $100$ different (equidistantly
distributed) starting positions for times up to $t=2^{22}-2^{28}$ and computed the
average displacement, $[\langle x(t) \rangle]_{\rm av}$. 
Besides, we have have considered finite systems of size 
$L=2^4-2^{14}$, $K=2$,  
generated $10^4-10^6$ independent random
environments for each system size and computed the average of 
$\ln\Omega_{\rm L}$ by the renormalization group method. 
Regarding the efficiency, the two methods complement each other.
The renormalization is more efficient in the vicinity of the
recurrent point, whereas by the simulation one can treat longer time-scales far from the recurrent point.

\begin{figure}[htb]
\begin{center}
\includegraphics[width=6.3cm,angle=0]{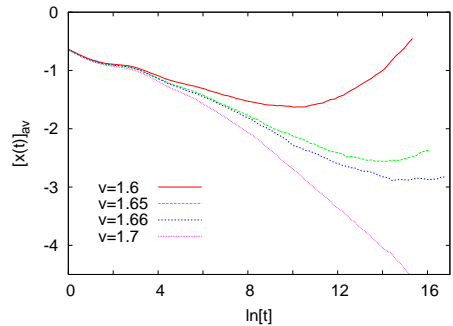}
\caption{Left: Average position (horizontal coordinate) of a walker in time for different values of $v$ for the bimodal
distribution in Eq.(\ref{bim_B}) with $c=0.4$ and $\lambda=0.1$ measured in 
Monte Carlo simulation. The direction of the motion is reversed at $v \approx 1.66$.}  
\label{fig3}
\end{center}
\end{figure}

We start with the asymmetric randomness, where we choose $c=0.4$ and $\lambda=0.1$ in the bimodal distribution in Eq.(\ref{bim_B}).    
The average displacements calculated by Monte Carlo simulations are shown in
Fig.\ref{fig3} for various values of the lane change rate $v$.

\begin{figure}[htb]
\begin{center}
\includegraphics[width=6.3cm,angle=0]{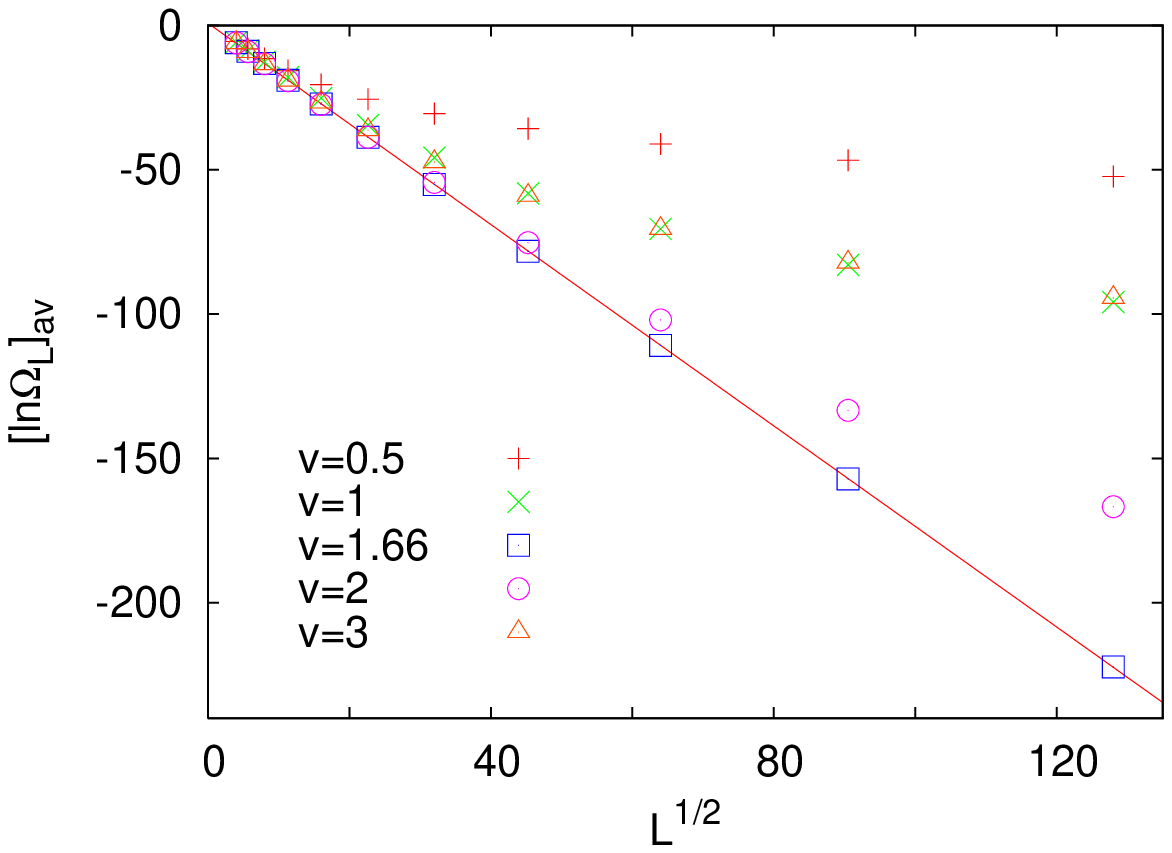}
\includegraphics[width=6.3cm,angle=0]{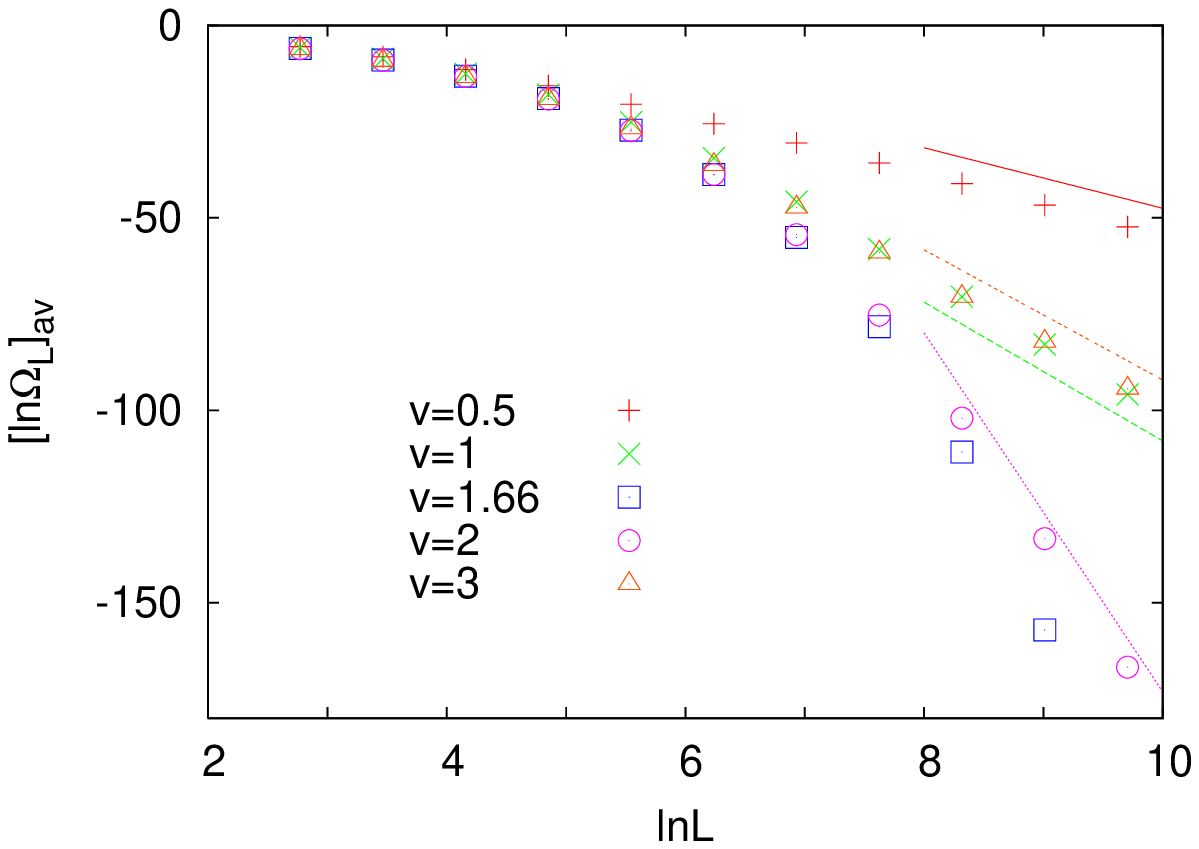}
\caption{The average of the logarithm of the minimal exit rate, 
$[\ln \Omega_L]_{\rm av}$, plotted against $L^{1/2}$ (left panel) and
against $\ln L$ (right panel) for the bimodal distribution used in Fig.\ref{fig3}.
At the recurrent point, $v=v_c$, the linear dependence in the plot in the left
panel corresponds to Sinai-scaling in Eq.(\ref{sinai}). 
In the transient phase, linear dependence is seen in the plot in 
the right panel, and the asymptotic values of the slopes
identified with  $1/\mu_2(v)$ depend on $v$ (see Eq.(\ref{JOmega})).}  
\label{fig4}
\end{center}
\end{figure}

As seen in this figure, the direction of the walk has changed by varying
$v$. For small lane change rate, $v<v_c\approx 1.66$
the walker goes ahead, whereas for large values of $v>v_c$, it moves
backwards. At $v=v_c$ the average displacement tends to a constant in the
limit $t\to\infty$. 
We have also studied the typical time-scale in the problem,
$t_L \sim \Omega_L^{-1}$, by the renormalization method. 
According to phenomenological scaling at the recurrent point we 
have $\ln\Omega_L\sim L^{1/2}$ (see Eq.(\ref{sinai})) 
whereas, in the transient phase, the dependence is algebraic: 
$\Omega_L \sim L^{-1/\mu_2}$ (see Eq.(\ref{JOmega})). 
In order to check both type of behavior, 
we have plotted $[\ln \Omega_L]_{\rm av}$ as a function of $L^{1/2}$ (Fig.\ref{fig4}, left panel) and as a function of
$\ln L$ (Fig.\ref{fig4}, right panel).

The results in Fig.\ref{fig4} are in complete agreement with the scaling
theory. In the transient phase, the diffusion exponent $\mu_2(v)$, which 
can be extracted from the asymptotic slopes 
of the curves in the right panel, are found to depend on $v$. 
We are going to study this issue in details for the symmetric
disorder, where the condition of recurrence is exactly known.

\begin{figure}[htb]
\begin{center}
\includegraphics[width=6.3cm,angle=0]{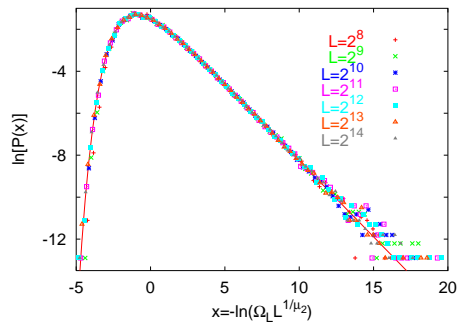}
\caption{Distribution of the scaling variable $\Omega_L L^{1/\mu_2}$, 
calculated for $K=2$ with symmetric bimodal randomness ($c=1/8,~\lambda^2=1/3$
and $v=0.1$). 
The scaling collapse is obtained with a diffusion exponent $\mu_2=0.726$. 
The solid line is the Fr\'echet-distribution given in Eq.(\ref{frechet}) 
with the parameter $u_0=0.392$.}  
\label{fig5}
\end{center}
\end{figure}

For symmetric disorder we consider also two-lane systems and use the bimodal randomness in Eq.(\ref{bim_A}).
In our first example we have $c=1/8$ and $\lambda^2=1/3$,
which, according to Eq.(\ref{mu_1_A}), results in a diffusion exponent 
$\mu_1=1/3$ for $K=1$. 
By application of the renormalization group procedure we have calculated sample dependent minimal
exit rates $\Omega_L$ and studied their distribution.
For different lengths, $L$, the appropriate scaling combination is: $\Omega_L L^{1/\mu_2}$,
see Eq.(\ref{JOmega}), and the distributions in terms of this reduced variable show a nice scaling collapse. This
is illustrated in Fig.\ref{fig5} for the lane change rate, $v=0.1$, using a diffusion exponent
$\mu_2=0.726$. The
scaling curve is very well described by the Fr\'echet-distribution\cite{galambos}:
\be
P_{Fr}(u)=\mu_2 u^{\mu_2-1} \exp(-u^{\mu_2})\;,
\label{frechet}
\ee
with $u=u_0 \Omega_L L^{1/\mu_2}$. Here the non-universal constant, $u_0$, which depends on the amplitude
of the tail of $P_>(\Omega)$ is the only free parameter. 

The distribution of the low-energy excitations in strongly disordered systems has been studied previously in Ref.\cite{extr}.
Using the strong disorder renormalization group method it has been argued that in several models the smallest
excitations, the energy of which follow an asymptotic power-law behavior, can be considered
asymptotically independent and therefore their distribution is in the Fr\'echet form. Our result in Fig.\ref{fig5}
indicates that the disordered multi-channel problem satisfies this conjecture.

\begin{figure}[htb]
\begin{center}
\includegraphics[width=6.3cm,angle=0]{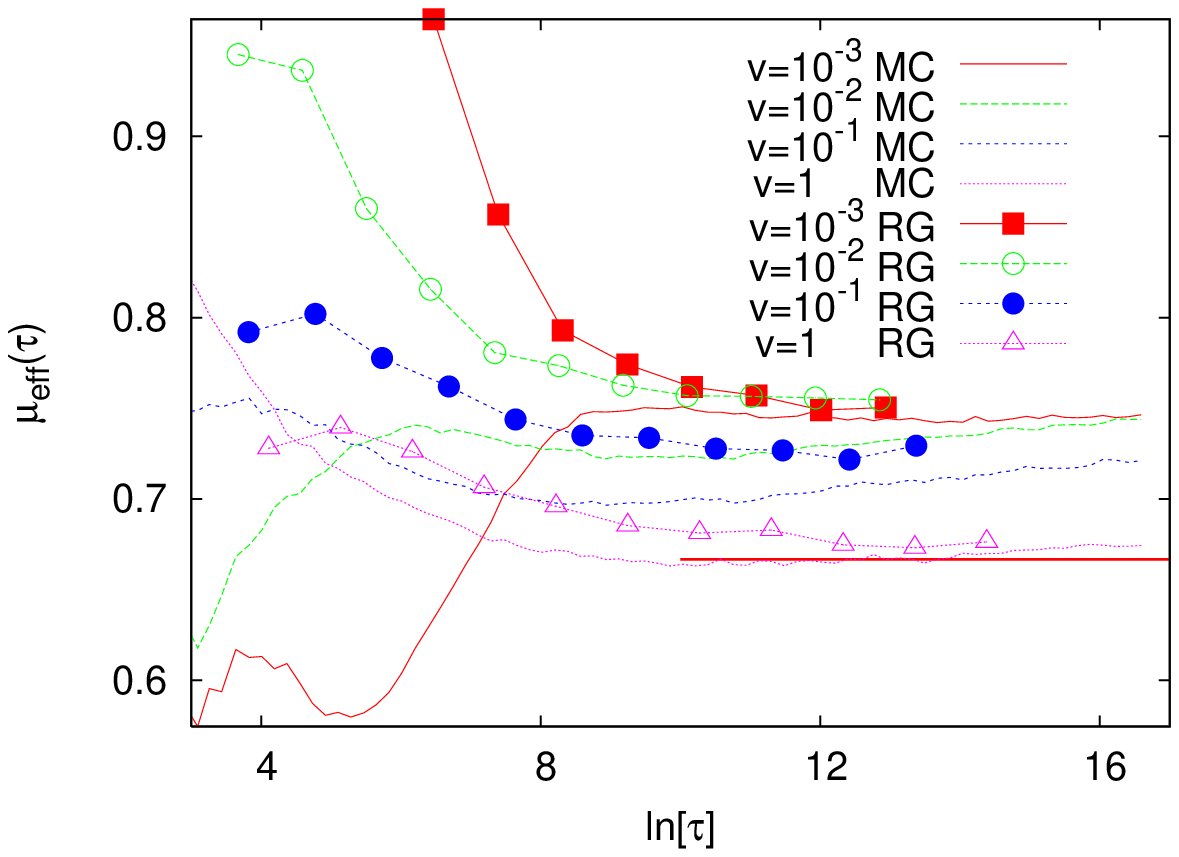}
\includegraphics[width=6.3cm,angle=0]{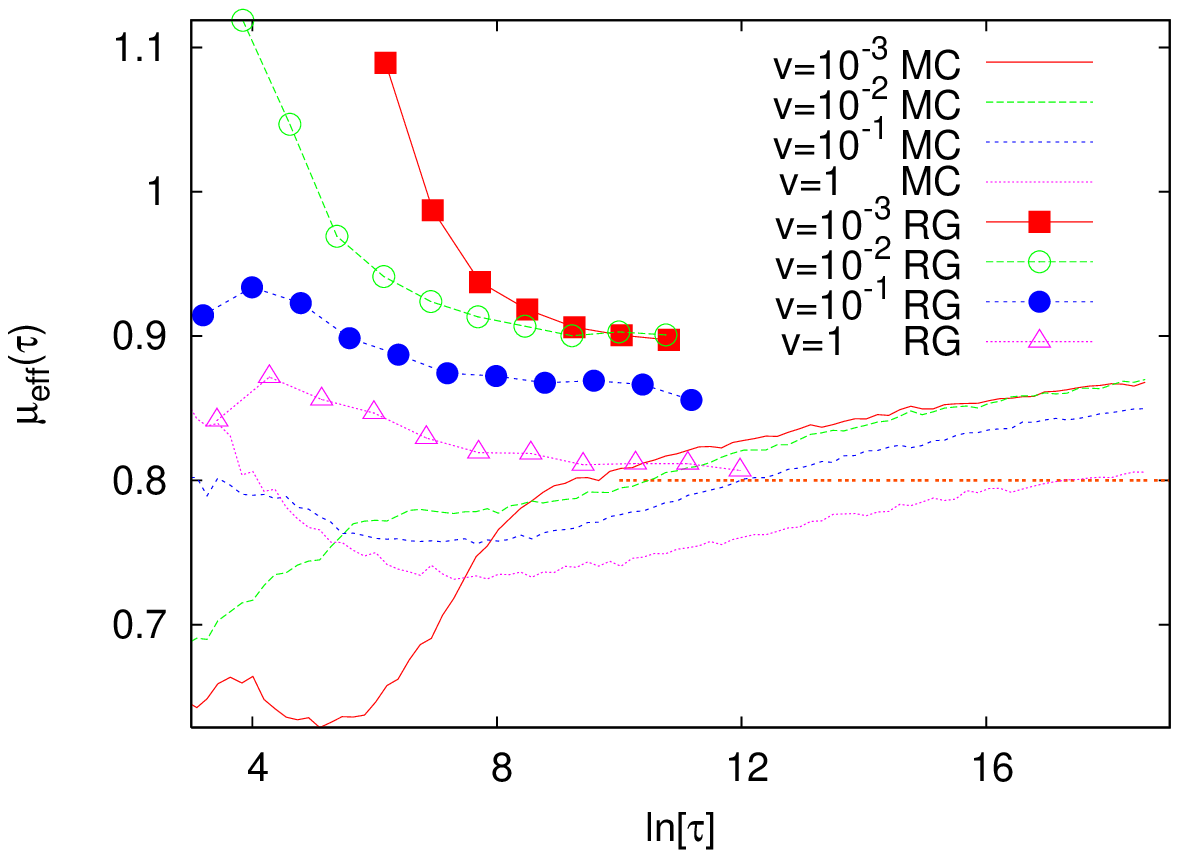}
\caption{Effective diffusion exponent of the random walk in the two-channel
problem for various values of the lane change rate, $v$, calculated either
by the Monte Carlo simulations ($\tau=t$) or by the renormalization method ($\tau=L^{\mu_2}$). We used symmetric bimodal randomness.
Left panel: $c=1/8,~\lambda^2=1/3$, so that $\mu_1=1/3$. Right panel: $c=1/3$,
$\lambda^2=0.1768$, so that $\mu_1=0.4$.}  
\label{fig6}
\end{center}
\end{figure}

We have repeated the previous renormalization group calculation for different values of $v$ and estimated the diffusion exponent by comparing
the average value of $[\ln\Omega_L]_{\rm av}$ and that of $[\ln\Omega_{L/2}]_{\rm av}$. From this we have obtained
effective exponents:
\be
\mu_2^{\rm eff}(L)=\frac{ [\ln\Omega_{L/2}]_{\rm av}-[\ln\Omega_{L}]_{\rm av}}{\ln 2}\;,
\ee
which are shown in the left panel of Fig.\ref{fig6} as a function of $\tau=L^{1/\mu_2}$. These exponents
are compared with those calculated by the Monte Carlo simulations. In this case the average position of the walker, $[x(t)]_{\rm av}$,
is calculated at time, $t$, and from the results at two different times, $t$ and $t\cdot\Delta$, we have
obtained:
\be
\mu_2^{\rm eff}(t)=\frac{\ln [x(t\cdot\Delta)]_{\rm av}-\ln [x(t)]_{\rm av}}{\ln\Delta}\;.
\ee
These are also shown in the left panel of Fig.\ref{fig6} as a function of $\tau=t$.

As can be seen in this figure the calculated effective exponents have strong $\tau$ dependence
for small systems (renormalization method) and for short times (simulation). 
However, the true exponents obtained by the two methods, which should be seen
in the limit $\tau\to\infty$ seem to agree well. 
As expected, $\mu_2$ is a continuous function of the lane change rate 
and it is found to decrease monotonically with $v$. 
In the large $v$-limit it approaches the known exact
limiting value: $\lim_{v \to \infty} \mu_2=2 \mu_1$, see Eq.(\ref{mu2}). Similar tendency is seen in the
right panel of Fig.\ref{fig6} for different parameters of the distribution:
$c=1/3$, $\lambda^2=0.1768$, so that $\mu_1=0.4$. In both distributions the diffusion
exponent remains smaller than one, even for very small value of $v$, thus the system
is always in the anomalous diffusion regime.

\begin{figure}[htb]
\begin{center}
\includegraphics[width=6.3cm,angle=0]{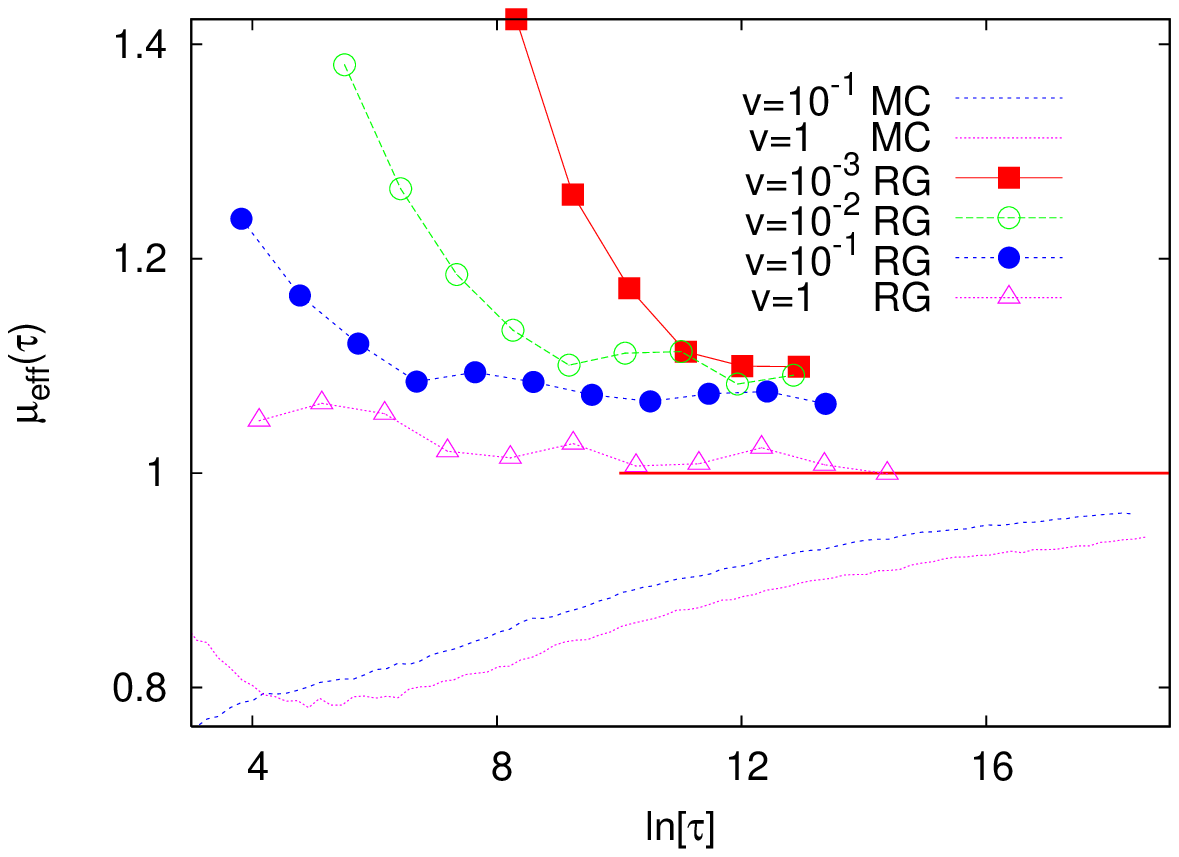}
\includegraphics[width=6.3cm,angle=0]{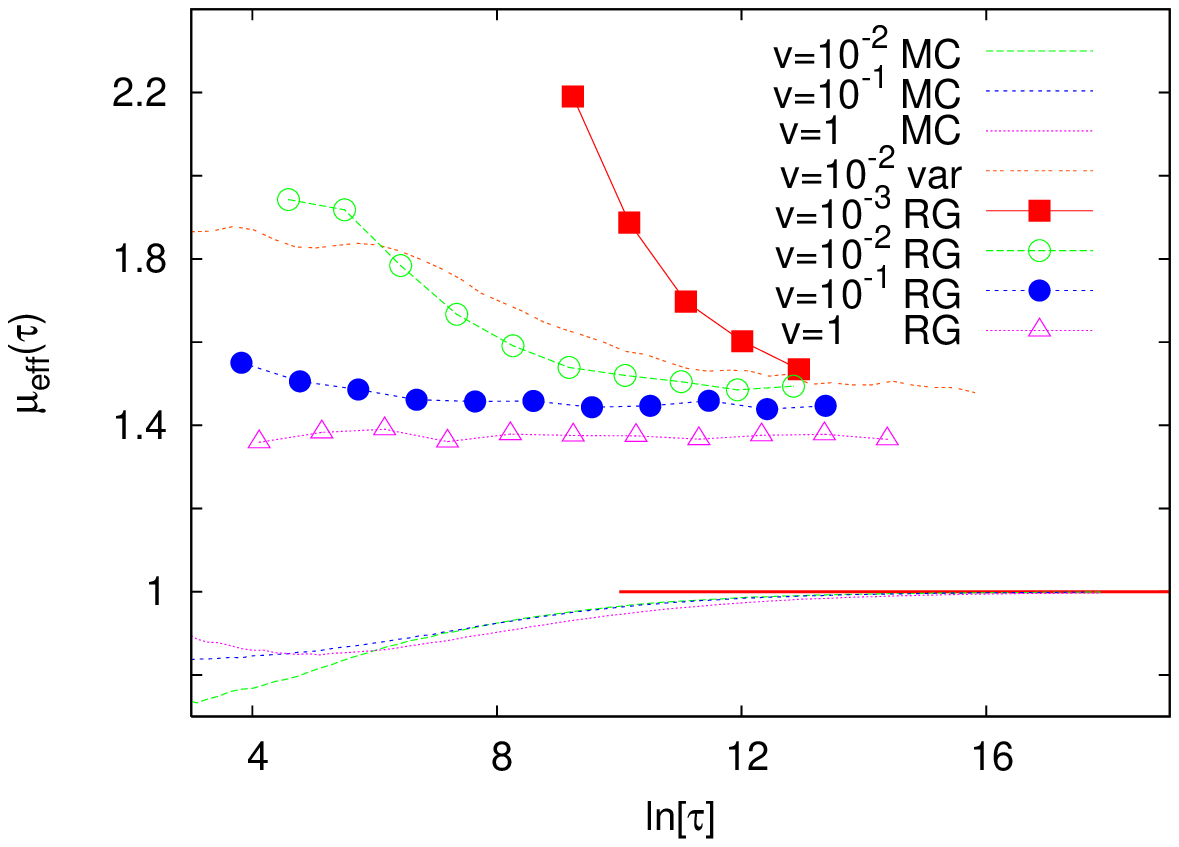}
\caption{The same as in Fig.\ref{fig6} for ballistic transport. Here the $\mu_{\rm eff}$ calculated
by the renormalization group method are the disorder induced diffusion exponents, which represent correction to scaling effects.
Left panel: $c=1/4$, $\lambda^2=1/9$,
so that $\mu_1=0.5$. Right panel: $c=1/4$, $\lambda^2=1/5$, thus $\mu_1=0.683$.
Here we also present the exponent $3-\sigma$ (indicated by "var"), in which $\sigma$ is the effective exponent of
the second moment of the displacement measured in the simulations.}  
\label{fig7}
\end{center}
\end{figure}

This type of behavior is going to change if we use such type of distributions, for
which $\mu_1$ reaches or exceeds the value of $0.5$. In this case $2 \mu_1 \ge 1$, thus the
transport in the two-lane system is ballistic. This type of behavior is illustrated in Fig.\ref{fig7} for $c=1/4$, $\lambda^2=1/9$,
i.e. $\mu_1=0.5$ (left panel) and for $c=1/4$, $\lambda^2=1/5$, i.e. $\mu_1=0.683$ (right panel).

Indeed the effective exponents calculated by the Monte Carlo method 
approach one, which is a clear
signal of the ballistic behavior. 
(We note that in the borderline case, $\mu_K=1$, there may be logarithmic
corrections to the ballistic behavior just as 
for $K=1$ at $\mu_1=1$ \cite{bouchaud})

In this case the exponent extracted from the finite size scaling 
of $\Omega_L$ is greater than $1$ and thus differs from the true 
diffusion exponent, which is equal to $1$. 
The exponent obtained by the renormalization in this phase 
appears in the corrections to scaling to the ballistic behavior. 
To be concrete, this exponent
governs the scaling of the second moment of the displacement, $\Sigma_L$,
which is expected to scale as $\Sigma_L \sim L^{\sigma}$,
with $\sigma=3-\mu_2$. In the right panel of Fig.\ref{fig7} the 
renormalization estimates for $\mu_2$ are
compared with the estimates for $3-\sigma$, which are obtained in Monte Carlo simulations and good agreement is found.

Finally, we mention that, for symmetric randomness, 
it is possible to drive the system from the sublinear
transient phase to the ballistic phase by varying the lane change rate.  

\section{Discussion}

In this paper we have studied the diffusion of a particle in a 
quasi-one-dimensional system which consists of $K$ disordered channels. 
We have investigated the transient phase by Monte Carlo simulations 
and by a variant of the strong disorder renormalization method.
In this phase, the diffusion exponent is found to depend on 
$K$ and the lane change rate.  
For asymmetric disorder, the recurrent point is shifted compared to that of
the one-lane model and the interaction between the lanes is able to cause 
a counter-intuitive reversal of the direction of motion when 
the lane change rate is varied. This phenomenon is analogous to the ratchet
effect in a saw-tooth potential.  
For symmetric disorder the recurrent point stays invariant and the diffusion
exponent in the transient phase is found to
increase with $K$ and decrease with the lane change rate.  

The numerical value of the diffusion exponent for $K>1$ and for intermediate
lane change rates $0<v<\infty$ is non-trivial. 
It is easy to see, that a naive random trap description, when applied
to the lanes separately, gives a wrong result. 
Indeed, identifying the trapping regions in each lane, then regarding them
point-like traps which are concentrated on a single site would lead after
simple argumentation on the probability of simultaneous occurrence of traps
with a waiting time at least $t$ in
all lanes to the diffusion exponent $\mu_K=K\mu_1$, independent of $v$ and the
type of randomness. 
The problem with this argument is that, although, the finite extension of
trapping regions does not play a role in one dimension,    
this circumstance cannot be disregarded in a coupled multi-channel system. 
After long times the
particle may encounter long enough overlapping trapping regions in all lanes, 
where it cannot pass even in the most favourable lane 
without changing lanes. 
Such overlapping regions must be treated as a whole and they lead
to a non-trivial dependence of the diffusion exponent on the distribution of
transition rates. 

The choice that the lane change rate is site-independent allows an 
alternative interpretation of the model.    
The particle can be thought to move in one-dimension but in a time-dependent
potential, which fluctuates stochastically with rate $v$ between 
$K$ different random potentials each of which is determined 
by the jump rates in the $K$ lanes. 
A similar but spatially non-random problem, the first passage 
time of a particle over a homogeneous barrier the height of which 
fluctuates between two different
values has been thoroughly studied \cite{doering,bier,lindenberg} 
In case of very low flipping frequency the mean first passage time is
dominated by first passage time over the higher barrier. 
For very high flipping frequency the particle feels an average potential 
and the first passage time is determined by the mean height of the potential. 
For intermediate frequencies the passing of the particle occurs primarily when
the height of the barrier is small.
Therefore, when varying the frequency, the mean first passage time exhibits a
minimum. This phenomenon is called resonant activation. 
The optimal frequency is a known to decrease exponentially with the minimal
height of the barrier \cite{lindenberg}.

Our result are qualitatively in agreement with the properties of resonant
activation. For long times, the particle encounters higher and higher
barriers, therefore the optimal lane change rate at which the release time is
minimal becomes smaller and smaller as the particle proceeds.
Consequently, for late times ($t\to\infty$), the diffusion exponent assumes
its maximum in the limit $v\to 0$, in agreement with the numerical results.
Note that these two limits can not be interchanged, therefore
the (asymptotic) diffusion exponent of the model as a function of 
the lane change rate is non-analytical at $v=0$.

\ack
This work has been
supported by the Hungarian National Research Fund under grant No OTKA
K62588, K75324 and K77629 and by a German-Hungarian exchange program (DFG-MTA).



\end{document}